\documentstyle[prl,aps,twocolumn]{revtex}

\draft

\author{M. A. Nielsen \thanks{Electronic address: mnielsen@tangelo.phys.unm.edu}}
\title{Computable functions, quantum measurements, and quantum dynamics}
\address{Center for Advanced Studies, Department of Physics and
Astronomy, University of New Mexico, Albuquerque NM 87131-1156 \\  {\em and} \\
Norman Bridge Laboratory of Physics 12-33, California Institute of
Technology, Pasadena, CA 91125}
\date{\today}

\begin{document}


\maketitle

\begin{abstract}
We construct quantum mechanical observables and unitary operators
which, if implemented in physical systems as measurements and dynamical
evolutions, would contradict the Church-Turing thesis which
lies at the foundation of computer science. We conclude that either
the Church-Turing thesis needs revision, or that only restricted classes
of observables may be realized, in principle, as measurements, and
that only restricted classes of unitary operators may be realized,
in principle, as dynamics.
\end{abstract}


\narrowtext

Quantum mechanical measurements on a physical system are represented
by {\em observables} - Hermitian operators on the state space of the
observed system.  It is an important question whether all observables
may be realized, in principle, as measurements on a physical system. Dirac's
influential text (\cite{Dirac58a}, page 37) makes the following assertion on the
question:

{\em The question now presents itself -- Can every observable be
measured? The answer theoretically is yes. In practice it may be very
awkward, or perhaps even beyond the ingenuity of the experimenter,
to devise an apparatus which could measure some particular
observable, but the theory always allows one to imagine that the
measurement can be made.}

This Letter re-examines the question of whether it is possible,
even in principle, to measure every quantum mechanical observable.
Unexpectedly, ideas from computer science are crucial
to the analysis. We also investigate a related question,
namely, whether it is possible to
realize, in principle, the dynamics corresponding to an arbitrary
unitary operator on the state space of a quantum system. Of course,
for specific systems particular measurements and unitary
dynamics may be forbidden by system specific features, such
as superselection rules. We will not be concerned with such specific
features, but rather with general considerations.

In the remarkable paper which founded modern computer science,
Turing \cite{Turing36a} defined a class of functions which are now known
as the {\em recursive} or {\em computable} functions. The
{\em Church-Turing thesis} \cite{Church36a,Turing36a} of computer
science states that this
class of functions corresponds precisely to the class of functions which
may be computed via what humans intuitively call an {\em algorithm}
or {\em procedure}. More formally,

{\em Every function which can be computed by what we would naturally
regard as an algorithm is a computable function, and vice versa.}

The Church-Turing thesis is fundamental to theoretical computer science, since it
asserts that the mathematical class of functions studied by computer scientists,
the computable functions, is the most general class of functions 
which may be calculated using a computer. It is an empirical
statement, not a theorem of mathematics, and has been
verified through over sixty years of testing \cite{Hofstadter79a,Penrose89a}.
For a review of different formulations of the Church-Turing thesis,
see \cite{Hofstadter79a}.

For convenience we refer to computable partial functions of a single
non-negative integer as {\em programs}. That programs are only partial
functions means that there may be some values of the input for
which no output is defined. An example of a program is the function
$f(x) = x^2$, which may be computed using a suitable Turing machine.
Using Turing's results it is possible to show \cite{Davis83a}
that the set of programs may be numbered $0,1,2,3,\ldots$. Not all programs
need to terminate or {\em halt} for all possible inputs. A simple example is the
program $f$ which on input $x$ loops forever if $x$ is not a perfect
square, or prints $\sqrt x$ if $x$ is a perfect square. This is an example
of a partial function.

The question of whether or not a given program, numbered $x$, halts
on the input of $y$ is obviously a question of considerable practical importance:
we would like to know whether a given algorithm will terminate or
not. To understand this question better, Turing defined a function,
the {\em halting function}, $h$, by
\begin{eqnarray}
h(x)  \equiv  \left\{ \begin{array}{ll} 1 & \mbox{if program} \, x \,
	\mbox{halts on input} \, x; \\
	0 & \mbox{if program} \, x\, \mbox{does not halt on input} \, x.
\end{array} \right.
\end{eqnarray}
Turing \cite{Turing36a} demonstrated that the halting function is
not a computable function. That is, there exists no algorithmic means
for computing the value of the halting function for all values $x = 0,1,\ldots$.
Thus there is no algorithm for determining whether a
given program terminates or not.

Many non-computable functions other than the halting function
are now known, and the reasoning which follows applies to any such function
$h$. For concreteness we continue to imagine that $h$ is
the halting function.

We define the {\em halting observable}, $\hat h$, by
\begin{eqnarray}
\hat h \equiv \sum_{x = 0}^{\infty} h(x) |x\rangle \langle x|, \end{eqnarray}
where $|x\rangle$ is an orthonormal basis for the state space
of some physical system with a countably infinite dimensional
state space. We will suppose that the system is one such that
all the states $|x\rangle$ may be prepared, in principle.
For example, the states $|x\rangle$ might be number states of
a single mode of the electromagnetic field.
The halting observable is clearly an observable in the
usual quantum mechanical sense: it is a Hermitian operator
on the state space of the system of interest.

Logically, one of two possibilities must hold:
\begin{enumerate}
\item It is possible, in principle, to construct a measuring device 
capable of performing a measurement of the observable $\hat h$.

\item It is not possible, in principle, to construct a measuring device 
capable of performing a measurement of the observable $\hat h$.
\end{enumerate}
Suppose the first possibility is true. Then in order to compute
the value of $h(x)$ one performs the following procedure:
Construct the measuring apparatus to measure $\hat h$, and prepare the
system to be measured in the state $|x\rangle$. Now perform the
measurement. With probability one the result of the measurement
will be $h(x)$. This gives a procedure for computing the
halting function. If one accepts the Church-Turing thesis this
is not an acceptable conclusion, since the halting function is not
computable.

Acceptance of the Church-Turing thesis therefore forces us to conclude 
that the second option is true, namely, that
it is not possible, in principle, to construct a measuring device capable of
performing a measurement of the observable $\hat h$. That is, only
a limited class of observables correspond to 
measurements which may be performed, in principle,
on quantum mechanical systems. An important question arises:
to determine the precise class of observables which may 
be realized as measurements.

Might it be possible to perform an approximate measurement of $\hat h$?
Suppose it is possible to measure an observable $\hat h'$ which
is close to $\hat h$. Preparing the system in the state $|x\rangle$ and
measuring $\hat h'$, a result in the range $h(x) \pm \delta$ is obtained with
probability at least $1-\epsilon$, for some small $\epsilon$ and
$\delta$. Clearly, by performing repeated measurements of this type 
it is possible to determine $h(x)$ with arbitrarily high confidence. Thus,
approximate measurements of $\hat h$ give an algorithmic
means for computing $h(x)$. Once again, if we accept the Church-Turing
thesis then we are forced to conclude that it is not
possible to perform such an approximate measurement.
Note, however, that we are implicitly using a stronger
version of the Church-Turing thesis than hitherto, since now we are regarding
as an algorithm a procedure which outputs $h(x)$ with {\em arbitrarily
high confidence}, rather than a purely deterministic procedure.

This second conclusion should be compared to work by Wigner \cite{Wigner52a},
and Araki and Yanase \cite{Araki60a} on the WAY theorem
\cite{Peres93a}. The WAY theorem shows that if we require certain
conservation laws to be respected during the measurement process, then
there are restrictions on the class of observables which may be realized
using measuring devices. This is fundamentally different to the
conclusion we have obtained, which does not depend on the imposition of
externally imposed conservation requirements. Another difference
is that it was shown in \cite{Araki60a} and \cite{Ghirardi81a} that it is
possible to perform approximate
measurements of the observables forbidden realization as measurements by 
the WAY theorem. As we have already seen, approximate
measurements of $\hat h$ are not allowed by the Church-Turing thesis.

Up to this point we have considered the physical realization of measurements
corresponding to quantum mechanical observables.
Similar arguments apply also to the physical realization of unitary
operators as dynamical evolutions. Define a function $g$ as follows:
\begin{eqnarray}
g(x) & \equiv & \left\{ \begin{array}{ll} 2m-2 & \mbox{if}\,x\,\mbox{is the}\,
  m\mbox{th smallest non-negative} \\
 & \mbox{ integer such that}\, h(x)=0 \\
	2m-1 & \mbox{if}\,x\,\mbox{is the}\,
  m\mbox{th smallest non-negative} \\
 & \mbox{integer such that}\, h(x)=1. \end{array}
 \right. \nonumber \\
 & & \end{eqnarray}
It is easy to verify that the operator
\begin{eqnarray}
U & \equiv & \sum_{x=0}^{\infty} |g(x)\rangle \langle x| \end{eqnarray}
is unitary. Suppose we prepare the system in the state
$|x\rangle$, perform the unitary dynamics $U$, and then do a
measurement in the $|x\rangle$ basis with outcome $x'$ (note that
there are systems where such a measurement
can certainly be done, in principle, such as a single mode of the
electromagnetic field). Note that
$x'$ is even if and only if $h(x)=0$ and $x'$ is odd if and only if $h(x)=1$,
so this gives a procedure for computing the halting
function.
Once again, logically, one of two possibilities must hold:
\begin{enumerate}
\item It is possible, in principle, to construct a system whose dynamics is
described by the unitary operator $U$.

\item It is not possible, in principle, to construct a system whose dynamics
is described by the unitary operator $U$.
\end{enumerate}
Once again, if we accept the Church-Turing thesis then we are forced
to the second conclusion: there are unitary operators which do not describe the
dynamics of any system which can, even in principle, be constructed.
By arguments similar to those used for observables, it is easy to see
that an approximate dynamical realization of $U$ can also
be used  as part of a procedure
for evaluating the halting function, so acceptance of the Church-Turing thesis
implies that approximate realizations of $U$ are not possible, either.

The examples we have discussed take place in infinite dimensional state spaces.
A similar construction for a spin $\frac 12$ system starts by defining (see chapter seven
of \cite{Cover91a} for a review of definitions along these lines, and references)
\begin{eqnarray}
\Omega \equiv \sum_{x: h(x)=1} \frac{1}{2^x}. \end{eqnarray}
Note that $0 < \Omega < 1$, and that the $x$th bit in the binary expansion
of $\Omega$ is one if and only if $h(x) = 1$, so knowing the binary expansion of
$\Omega$ is equivalent to knowing $h(x)$ for all $x$.
Define $U \equiv \exp(-i \Omega \sigma_y)$.
Starting the system in the $|\frac 12,\frac 12\rangle$ state (spin up in the $z$ direction) 
and applying the dynamics $U$ we see that the state after the dynamics is
\begin{eqnarray}
\cos (\Omega)  |\frac 12,\frac 12\rangle + \sin(\Omega) |\frac12,-\frac 12\rangle.
\end{eqnarray}
By repeatedly performing this procedure and making measurements
of $\sigma_z$ we may determine $\cos (\Omega)$ and
thus $\Omega$ to any desired accuracy, with arbitrarily high
confidence. It follows that we can determine the value of $h(x)$ for all
$x$. Once again, if we accept the Church-Turing thesis, then we are
forced to conclude that $U$ cannot be realized.
Notice that, unlike the earlier examples, this
procedure is not stable under perturbations of $U$. A slight change
in $U$ can result in an incorrect evaluation of $h(x)$.
Physically, uncontrolled interactions with the environment
will necessarily mean that $U$ is not implemented exactly,
and thus it is not possible to evaluate
the halting function using a dynamical realization
of $U$. Based on similar arguments it seems likely,
though I know of no rigorous general
proof, that any {\em finite dimensional} construction which allows
evaluation of a non-computable function is unstable against
perturbations, and therefore is not physically interesting.

Returning to the two physically interesting infinite dimensional
examples, what conclusions can be drawn? There are two
programs one might pursue.

The first program is to modify the Church-Turing thesis. Perhaps
there exist in nature quantum processes which can be used to
compute functions which are classically non-computable. It
is far-fetched, but not logically inconsistent, to imagine some
type of experiment - perhaps a scattering experiment - 
which can be used to evaluate the halting function.

Recognizing such a process poses some problems.
How could we verify that a process computes the 
halting function (or any other non-computable function)?
Because of the unsolvability of the halting
problem, it is not possible to verify directly that
the candidate ``halting process'' is, in fact,
computing the halting function.
Nevertheless, one can imagine inductively verifying
that the process computes the halting function. One would do
this by running a large number of programs on a computer for
a long time, and checking  that all the programs which halt are
predicted to halt by the candidate halting process, and that
programs predicted not to halt by the candidate
halting process have not halted.  Given sufficient
empirical evidence of this sort, one could then {\em postulate}
as a new physical law that the process computes
the halting function. 

What types of modification of the Church-Turing thesis
might be considered in this program? One approach is
to exclude quantum phenomena {\em by fiat} from the area of
application of the thesis. Approaches of this type have numerous 
problems. First, the boundary between quantum and classical
phenomena is rather fuzzy; where precisely does one draw the line?
Second, the approach is {\em ad hoc}; what motivates
the rejection of quantum phenomena from the area of
application of the Church-Turing thesis? Many other
modifications of the Church-Turing
might be attempted, however we will not discuss
them here, as no fully satisfactory modification has been found
by the author.

It is the author's conjecture that the Church-Turing thesis
is essentially correct, and that a more satisfactory program
is to address the problem of achieving a sharp characterization
of the class of observables and unitary dynamics which may
be realized in physical systems. At least two properties must
be satisfied by such a characterization:
\begin{enumerate}

\item It should be consistent with a (possibly sharpened) form of the
Church-Turing thesis.

\item  It should be clear that the measurements and dynamics contained
in that class are, in principle, realizable, and that all other measurements
and dynamics could never be realized, even in principle, in the
laboratory.

\end{enumerate}

How might one achieve such a characterization?
Deutsch \cite{Deutsch85a} has proposed what he
calls the Church-Turing {\em principle}, to distinguish it from
the less well formulated 
Church-Turing {\em thesis}. The statement of this principle reads:

{\em Every finitely realizable physical system can be perfectly simulated by a
universal model computing machine operating by finite means.}

Note that {\em finite means} here has the meaning that on any
{\em given} computation finite computational resources are
used. As for classical computers, unbounded resources are in principle
available, provided only finite resources are used on any given computation.
{\em Finitely realizable} is being used in the same sense as we have used
{\em realizable}: in principle, the system can be constructed in
a laboratory.

Given such a universal machine, the problem of studying what classes of
measurements and dynamics are realizable
is reduced to the study of properties of a single physical system, the
universal computing machine, since that system can be used to
simulate all other finitely realizable physical systems. This
reduction to studying the properties of a single system may make
it considerably easier to study the classes of measurements
and dynamics which are realizable.

For example, Deutsch showed that his proposed
universal computing machine can not
compute any function which is not computable on a
classical computer. Thus, his machine can not compute the
halting function. But, if we assume that the halting observable
can be measured using some physical system, then by
the Church-Turing principle the measurement could be simulated
on a universal computing machine, and the result of the
measurement determined. This contradicts Deutsch's assertion
that such a machine can not compute any
function which is not computable classically, and we conclude that
measurements of the halting observable are not possible, in
principle. Thus, if we accept that Deutsch's proposed
machine satisfies the Church-Turing principle then it follows
that the halting observable can not be measured.

What is gained by using arguments based on the Church-Turing
principle instead of arguments based on the Church-Turing thesis,
is that it may be possible to prove the Church-Turing principle
within known physical theory, for a
suitable universal model computing machine. Unfortunately, it
is not clear to this author that the theory of quantum computation
(see \cite{Ekert96a,DiVincenzo95a,Bennett95a} for a review),
which has developed from Deutsch's original
proposal, provides a candidate universal model computing machine. In
particular, it is not clear that the finite dimensional state
spaces accessed by quantum computers are sufficient to
simulate, with arbitrary accuracy, all the processes
one finds in nature. Natural processes may take place in
infinite dimensional state spaces, and it is has not been demonstrated
that all such processes can be well simulated
using a system with only a finite
number of state space dimensions. Regardless of whether
quantum computers satisfy the Church-Turing principle,
it is certainly the case that the specification of a universal
model computing machine satisfying the Church-Turing principle,
besides being important in its own right, would also
greatly simplify the question of characterizing the
classes of measurements and dynamics which are
realizable in physical systems.

This Letter has discussed two questions: what class of observables  
may be realized as quantum measurements; and what unitary operators
may be realized as quantum dynamics.
Using concepts from computer science we have constructed observables and
unitary operators whose physical implementation would
contradict the fundamental Church-Turing thesis of computer
science. We conclude that the introduction of new
concepts into computer science, physics, or both, is necessary to
resolve this contradiction.

\acknowledgements
I thank Tony Bracken, Ike Chuang, Phil Diamond, Hoi-Kwong Lo,
Gerard Milburn, John Preskill and Howard Wiseman for discussions about this paper.
This work began with the support of a 1993 summer
vacation scholarship in the University of Queensland Mathematics Department,
and was continued with the support of the Office of Naval Research
(Grant No. N00014-93-1-0116), the support of DARPA through
the Quantum Information and Computing (QUIC) institute
administered by the Army Research Office, and the Australian-American 
Educational Foundation (Fulbright Commission).


\end{document}